\documentclass[reprint,aps,showpacs,superscriptaddress,amsmath,amssymb,prl]{revtex4-1}
\usepackage{dcolumn}
\usepackage{bm}
\usepackage[dvips]{graphicx}
\usepackage{wrapfig,subfigure}
\usepackage{amssymb}
\usepackage{color}
\usepackage{ulem}

\newcommand{\xb}{{\bf x}}

\newcommand{\ad}{a^{\dagger}}

\begin{document}

\title{Spectrum of an oscillator with jumping frequency and the interference of partial susceptibilities}

\author{
M. I. Dykman}
\affiliation{Department of Physics and Astronomy, Michigan State University, East Lansing, MI 48824}
\author{M. Khasin}
\affiliation{Department of Physics and Astronomy, Michigan State University, East Lansing, MI 48824}
\author{J. Portman}
\affiliation{Department of Physics and Astronomy, Michigan State University, East Lansing, MI 48824}
\author{S. W. Shaw}
\affiliation{Department of Mechanical Engineering, Michigan State University, East Lansing, MI 48824}
\date{\today}

\begin{abstract}
We study an underdamped oscillator with shot-noise frequency fluctuations. The oscillator spectrum is determined by the interference of the susceptibilities for different eigenfrequencies. Depending on the parameters, it has a fine structure or displays a single asymmetric peak. For nano-mechanical resonators with a fluctuating number of attached molecules, the spectrum is found in a simple analytical form. The results bear on various types of systems where the reciprocal correlation time of frequency fluctuations can be comparable to the typical frequency jumps.
\end{abstract}


\maketitle

Oscillators with varying frequency are studied in many contexts. The frequency change underlies high-resolution mass sensing with nano-mechanical resonators, which is based on the change being proportional to the mass of a particle attached to the resonator \cite{Ekinci2004,*Naik2009,Cleland2005,Burg2007,Jensen2008}. It is also used in dynamic atomic force microscopy and in high resolution magnetic force microscopy \cite{Rugar2004,*Mamin2007,Kuehn2008,*Moore2009}. In quantum systems,
the Fock states of a vibrational mode of a trapped electron were detected from the frequency change of a nonlinearly coupled mode \cite{Peil1999}. Recently it was proposed to use such change for quantum measurements of mechanical shot noise in an optomechanical system \cite{Clerk2010a}.

The change of the oscillator frequency can be often thought of as a jump; it occurs over a much smaller time than the oscillator decay time or the typical inter-jump interval. In addition, the jumps are random. This is the case, e.g., where they result from attachment and detachment of molecules (or nano-particles) to a resonator \cite{Ekinci2004,Yong1989,Cleland2002} or from transitions between well-separated energy levels in a system coupled to the oscillator \cite{Peil1999,Clerk2010a,Ivanov1966a}. The oscillator dynamics is determined by the interrelation between the characteristic frequency change in a jump $\Delta$, jump rate $W$, and the oscillator decay rate $\Gamma$. Of utmost interest for classical and quantum measurements is the range where all these parameters are small compared to the oscillator eigenfrequency $\omega_0$ in the absence of jumps, $W,\Gamma,\Delta\ll \omega_0$.

In this paper we consider the susceptibility of an oscillator with respect to weak resonant driving and also the power spectrum of the oscillator. These characteristics are advantageous as they can be directly measured in the experiment. We find them for an arbitrary interrelation between the relevant parameters, including $\Gamma,\Delta$, and $W$. The sensitivity of the spectra makes it possible to use them for determining the parameters and the statistics of the underlying jump processes.

Random frequency jumps are noise \cite{Yong1989,Cleland2002}, they lead to spectral broadening. This is an important mechanism of dephasing \footnote{We are grateful to Y. Nakamura for an example of such dephasing mechanism in SQUID-based systems}, and the analysis below immediately extends to systems other than the oscillator. For oscillators, the spectral broadening due to other types of frequency fluctuations has been discussed in the literature, see \cite{Ivanov1966a,Elliott1965,Lindenberg1981,Gitterman_book2005} and references therein.

The spectral broadening should be qualitatively different depending on the ratio $\Delta/W$. For small and comparatively frequent jumps, where $\Delta/W\ll 1$, one can think of the jumps as causing diffusion of the oscillator phase. The resulting spectral broadening should be of the order of the phase diffusion coefficient $\sim\Delta^2/W$ \cite{Anderson1954}.

The region $\Delta \gtrsim W$ is more complicated. One can imagine the oscillator as having states with different eigenfrequencies $\omega_0+\Delta_N$ separated by $\sim \Delta$, where $N$ enumerates the states. The interstate jump rate is $\sim W$. Such states can be thought of as different realizations of the oscillator and should be distinguished from the Fock states in each realization. Naively, one might expect that the oscillator susceptibility $\chi(\omega)$ is a sum of independent partial susceptibilities $\chi(N;\omega)$ in states $N$. They would be proportional to the state populations $P_N$, and Im~$\chi(N;\omega)$ would have a form of Lorentzians centered at $\omega_0+\Delta_N$. However as shown below, this is the case only in the limit of large $\Delta/W$.

The inapplicability of the picture of independent partial susceptibilities can be understood by noticing that, in order to resolve frequencies separated by $\Delta$, one should measure the system for time $\gtrsim 1/\Delta$. Therefore for $W\gtrsim \Delta$, the frequencies $\omega_0+\Delta_N$ cannot be resolved. As we show, the susceptibility can still be formally described as a sum of partial susceptibilities, but the latter are no longer independent. Rather, the complex partial susceptibilities are coupled, their shape is strongly changed compared to the $\Delta/W\gg 1$ limit, and the overall susceptibility can be described as a result of their interference. This is somewhat similar to the physics underlying the paradox of the quantum harmonic oscillator \cite{Weisskopf1930,Zeldovich1969,Nitzan1973,DK_review84}.

We will consider an oscillator coupled to a thermal bath and driven by a resonant field $F\exp(-i\omega t)+$~c.c., with $|\omega-\omega_0|\ll \omega_0$. We will assume that the coupling to the bath is linear in the oscillator coordinate and weak, and that the density of states of the bath weighted with the interaction is smooth near $\omega_0$, so that the oscillator decay rate is essentially the same for all frequencies $\omega_0+\Delta_N,\omega$, cf. Ref.~\onlinecite{DK_review84}.

Frequency jumps will be considered as imposed externally, oscillator back action on the source of the jumps will be disregarded. This is a good approximation in many cases, molecule attachment-detachment and nonlinear coupling to an off-resonance mode being examples. Between the jumps and in the absence of the driving and coupling to the bath the oscillator is described by Hamiltonian $H_0=\hbar(\omega_0+\Delta_N)\ad a$ which almost stepwise varies in time; here $\ad$ and $a$ can be defined as raising and lowering operators of an oscillator with frequency $\omega$, if one disregards energy corrections $\propto (\omega_0+\Delta_N-\omega)^2/\omega$.

The driving field does cause transitions between different-$N$ states. The oscillator response to the field is thus determined by the matrix elements $\rho(N)$ of the density operator $\hat\rho$ that are diagonal with respect to $N$. Changing to the rotating frame with the canonical transformation $U(t)=\exp(-i\omega \ad at)$ and using the rotating wave approximation, we obtain the master equation
\begin{eqnarray}
\label{eq:master_general}
\dot \rho(N) &=& i(\delta\omega -\Delta_N)[a^{\dagger}a,\rho(N)]+i[F'a^{\dagger}+F'^*a,\rho(N)]\nonumber\\
&&- \hat\Gamma\rho(N) + \hat W\rho(N),\quad\delta\omega=\omega-\omega_0,
\end{eqnarray}
where $F'=F/(2M\hbar\omega)^{1/2}$ ($M$ is the oscillator mass). The operator $\hat\Gamma$ describes oscillator decay due to the coupling to the bath and has the standard form $\hat\Gamma\rho=\Gamma(\bar n+1)(\ad a\rho-2a\rho\ad + \rho\ad a)+\Gamma\bar n (a\ad\rho-2\ad\rho a + \rho a\ad)$, where $\bar n= [\exp(\hbar\omega_0/k_BT)-1]^{-1}$ is the Planck number.

The operator $\hat W$ describes transitions between states $N$ with different eigenfrequencies,
\begin{equation}
\label{eq:W_operator}
\hat W\rho(N)=\sum_r[ W(N-r;r)\rho(N-r)-W(N;r)\rho(N)]
\end{equation}
where $r$ enumerates the number of states over which the transition is made. This model describes, in particular, molecule attachment-detachment where molecules attach to a narrow region on the nano-resonator, so that the oscillator frequency is determined by the total number of attached molecules $N$. If molecules do not interact with each other, they attach/detach one by one, and
\begin{eqnarray}
\label{eq:model_W_attachment}
&&W(N;1)=WN_0,\qquad W(N;-1)=WN,\\
&&W(N;r)=0\quad {\rm for} \quad |r|>1; \qquad \Delta_N=-N\Delta.\nonumber
\end{eqnarray}
Here, $N_0$ is the mean number of attached molecules, which is determined by the externally controlled molecule flux; for molecules of mass $m_{\rm mol}$, $\Delta\propto m_{\rm mol}\omega_0/M > 0$. The velocity jump from a mass change is small for $m_{\rm mol}/M\ll 1$, it does not cause phase accumulation in time in contrast to the frequency change, and can be disregarded; this was also checked by simulations.

The linear response of the oscillator to the driving is characterized by the susceptibility ${\cal X}(\omega)$ which relates the mean oscillator coordinate to the driving force, $\langle q(t)\rangle = {\cal X}(\omega)F\exp(-i\omega t) + $~c.c. For $\omega$ close to $\omega_0$ we have ${\cal X}(\omega)=(\hbar/2M\omega)^{1/2}\langle a \rangle/F$, where the expectation value of operator $a$ is given by the stationary solution of Eq.~(\ref{eq:master_general}). Setting $\dot\rho(N)=0$ in Eq.~(\ref{eq:master_general}), multiplying this equation by $a$ and taking first the trace over the Fock states of the oscillator for a given $N$ (which we denote by ${\rm Tr}_0$ below) and then the trace over $N$, we obtain ${\cal X}(\omega)=(2M\omega)^{-1}\chi(\omega)$, where
\begin{eqnarray}
\label{eq:partial_susc_general}
&&\chi(\omega)=\sum_N \chi(N;\omega),\\
&&[\Gamma-i(\delta\omega -\Delta_N)]\chi(N;\omega)- \hat W\chi(N;\omega) = iP(N).\nonumber
\end{eqnarray}
Here, $\chi(N;\omega)={\rm Tr}_0a\rho(N)/F'$, whereas $P(N)={\rm Tr}_0\rho(N)$ is the stationary probability to find the oscillator in state $N$. From Eq.~(\ref{eq:master_general}), this probability is independent of the driving and dissipation and is given by equation
\begin{equation}
\label{eq:stationary_distribution}
\hat W P(N)=0, \qquad \sum\nolimits_NP(N)=1.
\end{equation}

Equations (\ref{eq:partial_susc_general}) describes the scaled susceptibility $\chi(\omega)$ as a sum of complex ``partial" susceptibilities $\chi(N;\omega)$ for each eigenfrequency state $N$. These susceptibilities are given by a set of linear equations. They are coupled to each other, and since they are complex and the phase relations are important, one can say that they interfere, with $\chi(\omega)$ determined by the result of this interference.

Analytical expressions for $\chi(\omega)$ can be obtained in the limiting cases. We start with the case where the frequencies in different states $N$ are strongly different, $|\Delta_N-\Delta_{N'}|\gg \Gamma, W$ for $N\neq N'$. Here, the partial spectra are almost independent from each other. To describe $\chi(N;\omega)$ near resonance, $\omega\approx \omega_0+\Delta_N$, one should keep only diagonal in $N$ terms in Eq.~(\ref{eq:partial_susc_general}), which gives
\begin{eqnarray}
\label{eq:fine_structure}
&&\chi(N;\omega)\approx iP(N)[\gamma_N-i(\delta\omega-\Delta_N)]^{-1}, \\
&&\gamma_N=\Gamma+\sum\nolimits_rW(N;r) \quad (|\Delta_N-\Delta_{N'\neq N}|\gg \gamma_N, |\delta\omega|).\nonumber
\end{eqnarray}
From Eq.~(\ref{eq:fine_structure}), Im~$\chi(N;\omega)$ has a Lorentzian peak at frequency $\omega_0+\Delta_N$. The area of the peak $\pi P(N)$ is determined by the population of state $N$. The halfwidth of the peak $\gamma_N$ depends on the oscillator decay rate $\Gamma$ and the total probability to switch from state $N$ to other states, which is given by the second term in Eq.~(\ref{eq:fine_structure}) for $\gamma_N$. The overall oscillator absorption spectrum Im~$\chi(\omega)$ has fine structure with peaks described by Eq.~(\ref{eq:fine_structure}).

In the opposite limit of small frequency change, $|\Delta_N|/\Gamma \to 0$, all partial spectra have the same shape,
\begin{equation}
\label{eq:single_Lorentzian}
\chi(N;\omega)=iP(N)(\Gamma- i\delta\omega)^{-1} \qquad (|\Delta_N|\ll \Gamma).
\end{equation}
The jump rate does not affect the solution in the limit $|\Delta_N|/\Gamma \to 0$. The spectrum as a whole is Lorentzian centered at frequency $\omega_0$. This resembles the paradox of the quantum harmonic oscillator. The susceptibility of the oscillator can be presented as a superposition of coupled partial susceptibilities corresponding to transitions between neighboring quantum levels. If the oscillator is nonlinear, the transitions occur at different frequencies and the corresponding spectral lines have different width, but for a linear oscillator, as a result of the interference, all partial spectra have the same shape \cite{DK_review84}.

The jump rate drops out of the susceptibility also for arbitrary $\Delta_N/\Gamma$ as long as $W\gg |\Delta_N|,\Gamma$. In this case the term $\hat W \chi(N;\omega)$ is the leading order term in Eq.~(\ref{eq:partial_susc_general}) for $\chi(N;\omega)$, and to first order in $W^{-1}$,
\begin{eqnarray}
\label{eq:large_W_zeroth_order}
&&\chi(\omega)=i[\gamma-i(\delta\omega-\bar\Delta)]^{-1},\qquad \bar\Delta=\sum\Delta_NP(N),\nonumber\\
&& \gamma=\Gamma+\sum P(M)\Delta_N\Delta_Mx_{\alpha}(N)\tilde x_{\alpha}(M)(-\lambda_{\alpha})^{-1},
\end{eqnarray}
where $\xb_{\alpha},\tilde\xb_{\alpha}$ and $\lambda_{\alpha}$ are the right and left eigenvectors and nonzero eigenvalues of matrix $\hat W$ (the stability of the oscillator stationary state implies $\lambda_{\alpha}<0$). From Eq.~(\ref{eq:large_W_zeroth_order}), Im~$\chi(\omega)$ is again a Lorentzian peak, but now centered at the average frequency $\omega_0+\bar\Delta$ and with halfwidth $\gamma$ that exceeds $\Gamma$ by $\sim \Delta^2/W$.

For an arbitrary relation between the oscillator parameters Eqs.~(\ref{eq:partial_susc_general}) can be easily solved numerically. An explicit analytical solution can be obtained for several models. A simple model is where frequency jumps result from nonlinear coupling to another oscillator (oscillator $B$), which is in thermal equilibrium; here $\Delta_N=N\Delta$ for small nonlinearity ($N$ is the energy level number of oscillator $B$). The effect of thermal transitions between neighboring levels of oscillator $B$ is described by Eqs.~(\ref{eq:master_general}), (\ref{eq:W_operator}) with $W(N;1)=W\bar n_B(N+1)$ and $W(N;-1)=W(\bar n_B+1)N$ ($\bar n_B$ is the Planck number of oscillator $B$). The power spectrum for the corresponding frequency jumps was analyzed in Ref.~\onlinecite{Dykman1973}, and the jumps as they occur in time were seen for an electron in a Penning trap \cite{Peil1999}.

One of the most interesting and important for applications is the model of molecule attachment-detachment Eq.~(\ref{eq:model_W_attachment}). To find $\chi(\omega)$ it is convenient to write $\chi(N,\omega)$ as a Fourier transform,
\begin{equation*}
\chi(N;\omega)=\int\nolimits_0^{\infty}dt e^{i\delta\omega\,t}\tilde\chi(N;t),\quad \tilde\chi(N;0)=iP(N).
\end{equation*}
Then Eq.~(\ref{eq:partial_susc_general}) becomes a set of homogeneous differential-difference equations for functions $\tilde\chi(N;t)$. It can be solved using the discrete Laplace transform method, i.e., changing to $x(z;t)=\sum_Nz^N\tilde\chi(N;t)$. This leads to a first-order linear partial differential equation for $x(z;t)$. Its solution immediately gives $\tilde\chi(t)=\sum_N\tilde\chi(N;t)$,
\begin{eqnarray}
\label{eq:time_representation}
&&\tilde\chi(t)=e^{-\Gamma t +iWN_0\xi t}\exp\left[N_0\xi^2\left(1-e^{-(W-i\Delta)t}\right)\right],\\
&&\chi(\omega)=\int\nolimits_0^{\infty} dt e^{i\delta\omega\,t}\tilde\chi(t); \qquad \xi=\Delta/(W-i\Delta).\nonumber
\end{eqnarray}

Equation (\ref{eq:time_representation}) gives the susceptibility of the oscillator with a jumping frequency as an integral of an elementary function. The shape of the susceptibility strongly depends on the interrelation between the frequency change per jump $\Delta$ and the jump rate $W$. The analysis of this shape can be conveniently done by rewriting Eq.~(\ref{eq:time_representation}) as
\begin{eqnarray}
\label{eq:chi_as_sum}
\chi(\omega)&=&\sum\nolimits_{k=0}^{\infty}\phi_k(\omega),\qquad \phi_k(\omega)=e^{N_0\xi^2}(-N_0\xi^2)^k/k!\nonumber\\
&&\times\left[\Gamma-i(\delta\omega + WN_0\xi) + k(W-i\Delta)\right]^{-1}.
\end{eqnarray}
Formally, Eq.~(\ref{eq:chi_as_sum}) for $\chi(\omega)$ looks like a sum of partial spectra. However, functions $\phi_k(\omega)$ differ from the partial susceptibilities $\chi(k;\omega)$ introduced earlier. They are close only in the limit of large frequency jumps, $|\Delta|\gg W$, as seen by comparing Eqs.~(\ref{eq:single_Lorentzian}) and (\ref{eq:chi_as_sum}). In this limit the spectrum Im~$\chi(\omega)$ is a set of well-separated equally spaced Lorentzian lines with halfwidths $\approx \Gamma+ (k+N_0)W$, for the model (\ref{eq:model_W_attachment}). Equation (\ref{eq:chi_as_sum}) describes also corrections to the lineshape due to finite $\Delta/W$.

Of interest is also the limit $N_0|\xi|^2\ll 1$ where either the frequency shift is small, $|\Delta|\ll W$, or the average number of attached molecules is small, $N_0\ll 1$. Here, the leading order term in $\chi(\omega)$ is $\phi_0(\omega)$. It gives a Lorentzian peak of Im~$\chi(\omega)$ centered at $\delta\omega= -(W^2/\Delta)N_0|\xi|^2$, with halfwidth $\gamma=\Gamma+ WN_0|\xi|^2$, consistent with Eq.~(\ref{eq:large_W_zeroth_order}) for $|\Delta|\ll W$. The frequency-jump induced broadening can be comparable with $\Gamma$ even for small $N_0|\xi|^2$ provided $\Gamma\ll W$. For large $W/|\Delta|$ we have $W|\xi|^2\propto 1/W$, the jump-induced broadening becomes weaker with increasing $W$ in agreement with the picture of motional narrowing \cite{Anderson1954}. The line shift $\delta\omega\approx -N_0\Delta$ is independent of $W$ for $W/|\Delta|\gg 1$ and linearly increases with $N_0$. Corrections $\sim N_0|\xi|^2$ lead to the onset of characteristic asymmetry of the spectrum Im~$\chi(\omega)$.

The evolution of the spectrum Im~$\chi(\omega)$ with varying parameters is illustrated in Figs.~\ref{fig:fine_structure} and {\ref{fig:smooth_spectrum}. Already for moderate frequency jumps, $\Delta/\Gamma=3$, the spectrum may display a well-pronounced fine structure, with the inter-peak distance $\approx \Delta$, as seen from Fig.~\ref{fig:fine_structure}~(a). With increasing jump frequency this structure is smoothed out, but the spectrum has a characteristic asymmetric shape, Fig.~\ref{fig:fine_structure}~(b). With further increase of $W/\Delta$ the spectrum becomes narrow; the effect of frequency jumps is clearly seen in the shift of the peak away from $\omega_0$.

\begin{figure}[h]
\includegraphics[scale=0.45]{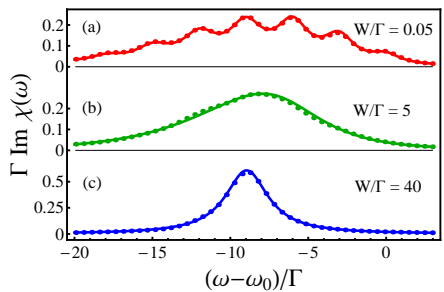}
  \caption{Color online. The spectrum of the oscillator with attaching and detaching molecules (nano-particles) for different scaled attachment rates $W/\Gamma$. The average number of attached molecules is $N_0=3$, and $\Delta/\Gamma=3$. Data points show the results of numerical simulations obtained using the Gillespie algorithm \cite{Gillespie1977}. }
  \label{fig:fine_structure}
\end{figure}

Figure \ref{fig:smooth_spectrum} demonstrates that, even for comparatively small $\Delta/\Gamma$, not only the position of the spectral peak, but also its shape are sensitive to the mean number of attached molecules $N_0$. This can be helpful in the analysis of experimental data, since $N_0$ can be changed by varying the influx of molecules.

\begin{figure}[h]
\includegraphics[scale=0.6]{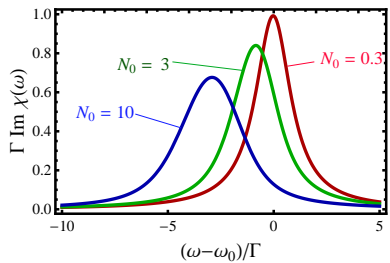}
  \caption{Color online. The evolution of the oscillator spectrum with the varying average number of attached molecules $N_0$ for a comparatively small frequency change per attached molecule. The parameters are $\Delta/\Gamma=0.3, W/\Gamma=0.1$.}
  \label{fig:smooth_spectrum}
\end{figure}

The results on the susceptibility give also the power spectrum of the oscillator in the absence of resonant driving $Q(\omega)=\pi^{-1}\,{\rm Re}\,\int_0^{\infty} dt\exp(i\omega t)\langle q(t)q(0)\rangle$. Function $Q(\omega)$ near its peak can be found from Eq.~(\ref{eq:master_general}) using the quantum regression theorem, which reduces the problem to equations of the form of Eq.~(\ref{eq:partial_susc_general}). The solution can be obtained directly or by thinking of the frequency fluctuations as having thermal origin, and then $Q(\omega)\approx (\bar n +1)(\hbar/2\pi M\omega)$~Im~$\chi(\omega)$ for $|\omega-\omega_0|\ll \omega_0$.

The above analysis can be extended to two-level systems with jumping inter-level spacing. If the occupation of the excited state can be neglected, the susceptibility is described by the set of linear equations (\ref{eq:partial_susc_general}). The problem becomes more complicated if the internal dynamics of the system that causes frequency jumps is important. An example is jumps due to nonresonant coupling to a driven cavity mode \cite{Gambetta2006,*Clerk2007,*Gambetta2008}. Here, partial susceptibilities of the system with jumping frequency have off-diagonal components with respect to the states of the system causing the jumps, hence instead of $\chi(N;\omega)$ one should consider $\chi(N,N';\omega)$. If the dynamics of the jump-causing system is Markovian in slow time and the system of interest is in the ground state, one obtains from the full master equation a set of linear equations for $\chi(N,N';\omega)$. This set is convenient for numerical analysis.

The results of this paper provide an insight into the spectral broadening from frequency jumps and a general method of describing it. They show that, by studying coupled partial susceptibilities, one can follow the evolution of the spectrum from well-resolved fine structure, for comparatively large and rare jumps, to a dephasing-type broadened single peak, for small and frequent jumps. The spectrum of a nanoresonator with attaching and detaching molecules or nano-particles is found in the explicit form. The susceptibility is sensitive to the parameters and the mechanism of the frequency jumps.

This research was supported in part by the NSF grants PHY-0555346 and CMMI-0900666 and ARO-57415-NS-II.

%

\end{document}